\documentclass[%
 aip,
 amsmath,amssymb,
 preprint,%
]{revtex4-1}

\usepackage{graphicx}
\usepackage{dcolumn}
\usepackage{bm}
\usepackage{xcolor}

\DeclareMathAlphabet{\mathcal}{OMS}{cmsy}{m}{n}

\usepackage[utf8]{inputenc}
\usepackage[T1]{fontenc}
\usepackage{mathptmx}
\usepackage[ruled,vlined]{algorithm2e}
\newcommand{\vx}{\boldsymbol{x}}
\newcommand{\vy}{\boldsymbol{y}}
\newcommand{\vp}{\boldsymbol{p}}
\newcommand{\vq}{\boldsymbol{q}}
\newcommand{\vf}{\boldsymbol{f}}
\newcommand{\vg}{\boldsymbol{g}}
\newcommand{\vQ}{\boldsymbol{Q}}
\newcommand{\vP}{\boldsymbol{P}}
\newcommand{\vm}{\boldsymbol{m}}
\newcommand*{\R}{\mathbb{R}}

\begin{document}


\title[Symplectic Gaussian Process Regression of Hamiltonian Flow Maps]{Symplectic Gaussian Process Regression of Hamiltonian Flow Maps}

\author{Katharina Rath}
\email{katharina.rath@ipp.mpg.de}
\affiliation{%
Department of Statistics, Ludwig-Maximilians-Universität München, Ludwigstr. 33, 80539 Munich, Germany
}%
\affiliation{ 
Max Planck Institute for Plasma Physics, Boltzmannstr. 2, 85748 Garching, Germany
}%

\author{Christopher G. Albert}%
\affiliation{ 
Max Planck Institute for Plasma Physics, Boltzmannstr. 2, 85748 Garching, Germany
}%

\author{Bernd Bischl}
\affiliation{%
Department of Statistics, Ludwig-Maximilians-Universität München, Ludwigstr. 33, 80539 Munich, Germany
}%

\author{Udo von Toussaint}
\affiliation{ 
Max Planck Institute for Plasma Physics, Boltzmannstr. 2, 85748 Garching, Germany
}%
\date{\today}

\begin{abstract}
We present an approach to construct appropriate and efficient emulators for Hamiltonian flow maps. Intended future applications are
long-term tracing of fast charged particles in accelerators and magnetic plasma confinement
configurations. The method is based on multi-output Gaussian process regression on scattered training data. To obtain long-term stability the symplectic property is enforced via the choice of the matrix-valued
covariance function. Based on earlier work on spline interpolation we observe derivatives of 
the generating function of a canonical transformation. 
A product kernel produces an accurate 
implicit method, whereas a sum kernel results in a fast explicit method from this approach. Both
correspond to a symplectic Euler method in terms of numerical integration. These methods are
applied to the pendulum and the Hénon-Heiles system and results compared
to an symmetric regression with orthogonal polynomials. 
In the limit of small
mapping times, the Hamiltonian function can be identified with a part of the generating function
and thereby learned from observed time-series data of the system's evolution. Besides comparable
performance of implicit kernel and spectral regression for symplectic maps, we demonstrate a substantial 
increase in performance for learning the Hamiltonian function compared to existing approaches.

\end{abstract}

\maketitle

\section{Introduction}
Many models of dynamical systems in physics and engineering can be cast into Hamiltonian form.
This includes systems with negligible dissipation found in classical mechanics, 
electrodynamics, continuum mechanics and plasma theory~\cite{Goldstein1980, Arnold1997, Marsden1999} 
as well as artificial systems created for numerical purposes such as Hybrid-Monte-Carlo 
algorithms~\cite{Neil2011} for sampling from probability distributions. 
A specific feature of Hamiltonian systems is their long-term behavior with conservation of 
invariants of motion and lack of attractors to which different initial conditions converge. 
Instead a diverse spectrum of resonant and stochastic features emerges that has been 
extensively studied in the field of chaos theory~\cite{lichtenberg1992}. These particular 
properties are a consequence of the symplectic structure of phase space together with 
equations of motion based on derivatives of a scalar field -- the Hamiltonian $H$.

Numerical methods that partially or fully preserve this structure in a discretized system
are known as geometric or symplectic integrators~\cite{Hairer2006}. Most importantly
such integrators do not accumulate energy or momentum and remain long-term stable at
relatively large time-steps compared to non-geometric methods. Symplectic integrators
are generally (semi-)implicit and formulated as (partitioned) Runge-Kutta schemes
that evaluate derivatives of $H$ at different points in time.

The goal of this work follows a track to realize even larger time-steps by interpolating the 
flow map~\cite{Abdullaev2006, Berg1994, Kasilov1997, Kasilov2002, warnock2009}
describing the system's evolution over non-infinitesimal times. For this purpose some 
representative orbits are integrated analytically or numerically for a certain period of time. Then the 
map between initial and final state of the system is approximated in a functional basis. 
Once the map is learned it can be applied to different states to traverse time in ``giant'' steps.
Depending on the application this can substantially reduce computation time.
When applied to data from measurements this technique allows to learn the dynamics,
i.e. the Hamiltonian of a system under investigation~\cite{bertalan2019}. 

\begin{figure}
\begin{minipage}[t]{0.49\linewidth}
    \includegraphics[width=\linewidth]{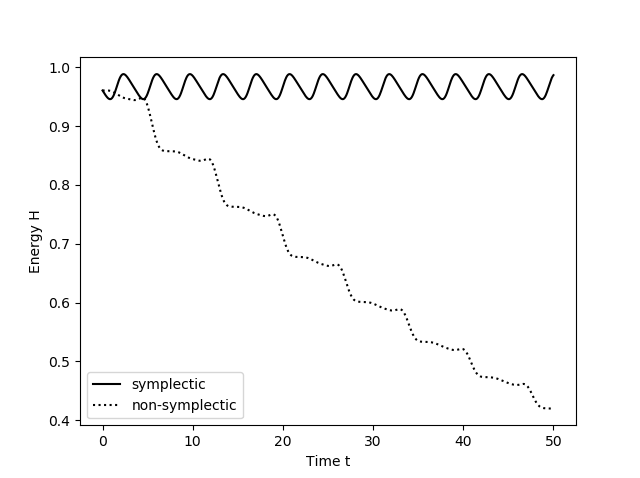}
\end{minipage}
\begin{minipage}[t]{0.49\linewidth}
    \includegraphics[width=\linewidth]{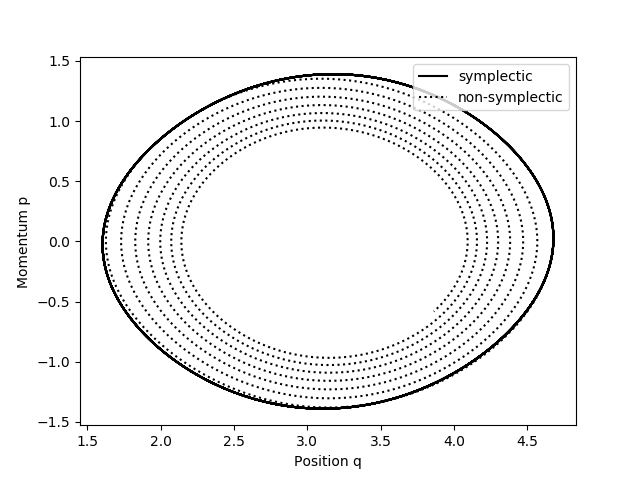}
\end{minipage}
\caption{Illustration of energy preservation and orbits in phase space using symplectic and non-symplectic interpolation of a pendulum's flow map}
\label{fig:symplnonsympl}
\end{figure}

Naively interpolating a map in both, position and momentum variables destroys 
the symplectic property of the Hamiltonian flow. In turn, all favorable properties
of symplectic integrators are lost and subsequent applications of the map 
become unstable very quickly. This problem is illustrated in Fig. \ref{fig:symplnonsympl},
where the flow map of a pendulum is interpolated in a symplectic and a non-symplectic manner,
respectively.
If one enforces symplecticity of the
interpolated map by some means, structure-preservation and long-term stability
are again natural features of the approximate map. Here this will be realized via
generating functions introduced by~Warnock~et~al.~\cite{Berg1994, warnock2009} in 
this context. This existing work relies on a tensor-product basis of Fourier series 
and/or piecewise spline polynomials. This choice of basis has two major drawbacks:
rapid decrease of efficiency in higher dimensions and limitation to box-shaped domains.
One possibility to overcome these limitations would be the application of 
artificial neural networks with symplectic properties\cite{greydanus2019hamiltonian, chen2019symplectic,burby2020fast,toth2019hamiltonian}. 
Here we rather introduce a kernel-based method as a new way to construct 
approximate symplectic maps via Gaussian process (GP) regression and radial basis functions
(RBFs). In the results it will become apparent
that such a method can work with much less
required training data in the presented test cases.

GP regression~\cite{rasmussen2005}, also known as Kriging, is a 
flexible method to represent smooth functions based on covariance functions (kernels) 
with tunable parameters. These kernel hyperparameters can be directly optimized in the training process by maximizing the marginal likelihood.
Predictions, in particular posterior mean 
and (co-)variance for function values, 
are then made via the inverse kernel covariance matrix. Observation of derivatives required
to fit Hamiltonian flow maps is possible via correlated multi-output Gaussian processes \cite{solak2003,eriksson2018, ohagan1992, alvarez2012}. During the construction, the close
relation to (symmetric) linear regression\cite{seber2012linear} and non-symmetric mesh-free collocation methods will 
become apparent~\cite{fasshauer1996,kansa1990}. 

The paper is structured as follows: First, Hamiltonian systems and canonical transformations that preserve the symplectic structure of phase space are introduced. Then, general derivations of multi-output Gaussian processes with derivative observations in the context of symmetric meshless collocation and of non-symmetric collocation with radial basis functions or orthogonal polynomials is given followed by a presentation of two algorithms to construct and apply symplectic mappings using Gaussian processes. Finally, the presented methods are tested on a simple pendulum and the more complex Hénon–Heiles system and compared to non-symmetric
collocation using radial basis functions as well as linear regression using an expansion in Hermite polynomials combined with a periodic Fourier basis.



\section{Dynamical Hamiltonian systems and symplectic flow maps}

Hamiltonian mechanics describe motion of a dynamical system in phase space, that has the structure of a symplectic manifold \cite{Arnold1997}. In the Hamiltonian formulation the equations of motion are of first order, so trajectories in phase space can never intersect
each other. For perturbation theory and to understand the general character of motion, the Hamiltonian point of view can yield important insights and keep the underlying 
symplectic phase space structure. 

\subsection{Hamiltonian systems}
A $f$-dimensional classical mechanical system is fully characterized by its Hamiltonian function $H(\vq, \vp, t)$, which is a function of $f$ generalised coordinates $\vq$, $f$ generalised momenta $\vp$ and time $t$. The time evolution of an orbit $(\vq(t),\vp(t))$ in phase space is given by
Hamilton's canonical equations of motion
\begin{align}
    \dot{\vq}(t) &= \frac{d\vq(t)}{dt}  = \frac{\partial H(\vq(t), \vp(t))}{\partial \vp}, \\
    \dot{\vp}(t) &= \frac{d\vp(t)}{dt}  = - \frac{\partial H(\vq(t), \vp(t)) }{\partial \vq}. 
    \label{eq:eom}
\end{align}

The Hamiltonian flow $\varphi_H$ intuitively describes the map that transports the collection
of all phase points along their respective orbits that are uniquely defined by initial
conditions.
More precisely, the derivatives of $H$ define a Hamiltonian vector field 
$\textbf{X}_H (\vq, \vp)$ on the cotangent bundle $\mathbb{T}^* \mathbb{Q}$ of
configuration space \cite{Arnold1997},
    \begin{equation}
    \boldsymbol{X}_H (\vq, \vp) =
    \left(\begin{array}{c}
    \nabla_{\vp} H(\vq, \vp) \\
    -\nabla_{\vq} H(\vq, \vp) \\
    \end{array}\right) = J^{-1}\nabla_{\vq, \vp}H(\vq, \vp),
    \label{eq:Xham}
    \end{equation}
    where the Poisson tensor in canonical representation
    \begin{equation*}
    J^{-1} \equiv \left(\begin{array}{cc}
    0 & I\\
    -I & 0 
    \end{array}\right)
    \end{equation*}
    is an $2f \times 2f$ antisymmetric block matrix with $I$ the $f \times f$ unit diagonal matrix. 
    Evolving a system along $\textbf{X}_H (\vq, \vp)$ over finite time intervals yields the Hamiltonian flow map $\varphi_H$, and preserves the symplectic structure of phase space \cite{Arnold1997}. The integral curves of $\textbf{X}_H (\vq, \vp)$ are solutions to the equations of motion given in Eq. \ref{eq:eom}. 
Important properties follow directly from the preservation of the symplectic structure of $\mathbb{T}^* \mathbb{Q}$: \textit{conservation} of invariants such as energy and momentum, and \textit{volume preservation} (as proven by Liouville's theorem) in phase space \cite{Arnold1997}. The latter means, if some region $\Omega$ is evolved according to the Hamiltonian flow $\varphi_H$, the volume of $\varphi_H(\Omega)$ 
remains constant. In a differential sense this means that the flow in phase space is
divergence-free:
\begin{equation}
    \nabla_{\vq} \cdot \nabla_{\vp} H - \nabla_{\vp} \cdot \nabla_{\vq} H = 0.
\end{equation}

\subsection{Canonical transformations}
One is usually interested in the temporal evolution according to Eq. \ref{eq:Xham}, that is, position $\vQ$ and momentum $\vP$ of a system at time $t=t_1$ that has been initialized with position $\vq$ and momentum $\vp$ at time $t=t_0$. Motion (or a shift in time) in an Hamiltonian system corresponds to a canonical transformation that preserves the form of the Hamiltonian vector field $\boldsymbol{X}_H$, and thereby invariants of a perturbed Hamiltonian \cite{Hairer2006} and the phase space volume. Also the canonical equations hold for the transformed coordinates $(\vQ, \vP)$. A common analytical technique to integrate Hamilton's canonical equations uses generating functions~\cite{Goldstein1980}. Generating functions are
also used as a way to construct symplectic integration schemes~\cite{Hairer2006}. Due to the Hamiltonian structure of equations of motion the mapping relations linking $\vq,\,\vp,\,\vQ$ and $\vP$ are not independent from each other, but linked via the symplectic property
\begin{equation}
    \frac{\partial \vQ(\vq,\vP)}{\partial \vq} - \frac{\partial \vp(\vq,\vP)}{\partial \vP} = 0.
    \label{eq:sympl}
\end{equation}
This property is closely related to divergence- or curl-freeness of vector fields. Similar to using a scalar or vector potential to guarantee such properties, 
symplecticity~(Eq. \ref{eq:sympl}) can be automatically fulfilled by introducing a generating function  $F(\vq, \vP)$. This function links old coordinates $(\vq, \vp)$ to new coordinates $(\vQ, \vP)$ via a canonical transformation. For a type 2 generating function \cite{Goldstein1980} this canonical transformation is given by
\begin{align}
    \vQ(\vq,\vP) &= \frac{\partial F(\vq,\vP)}{\partial \vP}, \\
    \vp(\vq,\vP) &= \frac{\partial F(\vq,\vP)}{\partial \vq}.
\end{align}
As the kernel regression of a linear term is less favorable, the generating function that maps several timesteps in the evolution of the Hamiltonian system is split into a sum,
\begin{equation}
    F(\vq, \vP) = \vq\cdot\vP + \Tilde{F}(\vq, \vP).\label{eq:cansplit}
\end{equation}
It's easy to check that the first part $\vq\cdot\vP$ in Eq.~\ref{eq:cansplit} describes the 
identity transformation $\vq\rightarrow\vQ,\,\vp\rightarrow\vP$.
The relation between $(\vq, \vp)$ and $(\vQ, \vP)$ can be written as
\begin{equation}
    \begin{pmatrix} \frac{\partial}{\partial \vq} \\ \frac{\partial}{\partial \vP} \end{pmatrix} \Tilde{F}(\vq, \vP) = \begin{pmatrix} \vp(\vq, \vP) - \vP \\ \vQ(\vq, \vP) - \vq \end{pmatrix} = \begin{pmatrix} -\Delta \vp(\vq, \vP)\\ \Delta \vq(\vq, \vP) \end{pmatrix}.
    \label{eq:grad}
\end{equation}
As any transformation via a generating function is canonical per definition \cite{Goldstein1980}, also the mapping created using a generating function preserves the symplectic structure of phase space. 

In the limit of small mapping times, the Hamiltonian $H$ can be identified (up to a constant) with the (differential) generating function $\Tilde{F}$, as time evolution is considered to be an infinitesimal canonical transformation. Namely, from Eq.~\ref{eq:grad}, we obtain the following expressions for $(\vQ, \vP)$:
\begin{align}
    \vQ &= \vq + \frac{\partial \tilde{F}}{\partial \vP}, \\
    \vP &= \vp - \frac{\partial \tilde{F}}{\partial \vq}. 
\end{align}
This can be compared by the equations of motion in Eq.~\ref{eq:eom}, where the first order approximation yields a symplectic Euler integration step
\begin{align}
    \vQ &\approx \vq + \Delta t \frac{\partial H}{\partial \vP}, \\
    \vP &\approx \vp - \Delta t \frac{\partial H}{\partial \vq}.
\end{align}
Comparing those sets of equations yields the relation $\tilde{F} = H \Delta t $ up to an irrelevant constant shift, where $\Delta t = t_1 - t_0$ is the mapping time. 
%
%

\section{Regression of Hamiltonian flow maps}
\subsection{Derivative observation in non-symmetric collocation}
Collocation and regression via basis functions approximates observed function values $g(\vx) \in \mathbb{R}^D$ for $\vx \in \R^d$  by fitting a linear combination of the chosen basis $\varphi_i(\vx)$,
\begin{equation}
g(\vx) = \sum_{i = 1}^n \alpha_i \varphi_i(\vx), 
\label{eq:rbf}
\end{equation}
where $\alpha_i$ are the weights and $n$ is the number of basis functions. Suitable bases are e.g. orthogonal polynomials,
splines, trigonometric functions (Fourier series) or radial basis functions with
kernels $\varphi_i(\vx) \equiv \varphi(\vx, \vx_i)$. In order to 
compute the weights $\alpha_i$, we solve
\begin{equation}
    \boldsymbol{\Phi} \boldsymbol{\alpha} = g(X), 
\end{equation}
where $\boldsymbol{\Phi} \in \R^{N \times n}\colon \Phi_{ij} = \varphi_j(\vx_i) $ for $N$ training points and  $X$ is the $d \times N$ design matrix. The interpolant $g_{*}$ at any point $\vx_{*}$ is given by
\begin{equation}
    {g}_{*}(\vx_{*}) = \sum_{i = 1}^n \alpha_i \varphi_i(\vx_{*}).
\end{equation}
When dealing with derivative observations, applying a linear operator $\mathcal{L}$, e.g. differentiation, to Eq.~\ref{eq:rbf} yields
\begin{equation}
    \mathcal{L}g(\vx) = \sum_{i=1}^n \alpha_i \mathcal{L}\varphi_i(\vx), 
    \label{eq:rbfdiff}
\end{equation}
which, when combining $g(\vx)$ with $\mathcal{L}g(\vx)$, results in the collocation matrix 
\begin{equation}\label{eq:PhiUnsym}
    \Tilde{\boldsymbol{\Phi}} = \begin{pmatrix}
    \boldsymbol{\Phi} \\
    \boldsymbol{\Psi}
    \end{pmatrix},
\end{equation}
with $\boldsymbol{\Psi}_{ij} = \mathcal{L} \varphi_j(\vx_i)$. In case of radial basis functions (RBFs), $\mathcal{L}$ is 
applied on the kernel function $\varphi(\vx_i, \vx_j)$ only once in the first argument.
The resulting linear system is usually overdetermined and has to be solved in a least-squares sense.
When applied to partial  differential equations the resulting method is called \textit{non-symmetric} 
or \textit{Kansa's method}~\cite{kansa1990, fasshauer1996}. 

\subsection{Derivative observation in (symmetric) linear regression}

In order to obtain a directly invertible positive definite system, one may instead use a
\textit{symmetric} least-squares regression method in a product basis\cite{seber2012linear}. Multiplying Eq.~\ref{eq:rbf} by $\varphi_{j}(x_k)$,
\begin{align}
g(\vx)\varphi_{i}(\vx) = \sum_{j = 1}^n \varphi_i(\vx)\varphi_{j}(\vx) \alpha_j, 
\end{align}
and subsequently summing over $N$ observations, 
\begin{align}
\sum_{k=1}^N \vy_k\varphi_{i}(\vx_k) = \sum_{k=1}^N \sum_{j = 1}^n \varphi_i(\vx_k)\varphi_{j}(\vx_k)\alpha_j 
\end{align}
we arrive at a minimization problem corresponding to $\boldsymbol{A} \boldsymbol{x}=\boldsymbol{b}$, with
\begin{align}
A_{ij} &= \sum_{k=1}^N \varphi_{i}(x_k) \varphi_{j}(x_k), \\
b_{i} &= \sum_{k=1}^N \varphi_{i}(x_k) \boldsymbol{y}_k.
\end{align}


When derivative observations are considered, the basis changes to $\psi_{i}=\mathcal{L}\varphi_i$, resulting in a minimization problem $\boldsymbol{A} \boldsymbol{x}=\boldsymbol{b}$, with
\begin{align}
A_{ij} &= \sum_{k=1}^N  \psi_{i}(x_k) \cdot \psi_{j}(x_k), \\
b_{i} &= \sum_{k=1}^N  \psi_{i}(x_k) \cdot \boldsymbol{y}_k.
\end{align}

Here, $\boldsymbol A$ is a symmetric positive definite matrix, that is directly invertible. 
%
%

\subsection{Multi-output GPs and derivative observations}

A Gaussian process (GP) \cite{rasmussen2005} is a stochastic process with the convenient property that any finite marginal distribution of the GP is Gaussian. 
For 
$\vx \in \mathbb{R}^{d} $, a GP with mean $\boldsymbol{m}(\vx)$ and kernel or covariance function $K(\vx, \vx^\prime)$ is denoted as
\begin{equation}
    \boldsymbol{f}(\vx) \sim \mathcal{GP}(\boldsymbol{m}(\vx), K(\vx, \vx^\prime)), 
\end{equation}
where we allow vector-valued functions~\cite{alvarez2012}. In contrast to a single output case, where the random variables are associated to a single process for $f(\vx) \in \mathbb{R}$,
a multi-output GP for $\boldsymbol{f}(\vx) \in \mathbb{R}^D$ consists of random variables associated to different and generally correlated processes. 
The covariance function is a positive semidefinite matrix-valued function, whose entries $(K(\vx, \vx^\prime))_{ij}$ express the covariance between the output 
dimensions $i$ and $j$ of $\boldsymbol{f}(\vx)$. In case a linear model for the mean $\boldsymbol{m}$ with some functional basis $\varphi_i$ and unknown
coefficients is used, a modified Gaussian process follows according to Rasmussen\&Williams\cite{rasmussen2005}, chapter 2.7.

For regression via a GP we assume that the observed function values $Y \in \mathbb{R}^{D \times N}$ may contain local Gaussian noise $\epsilon$ with covariance matrix $\Sigma_n$, i.e. the noise is independent
at different position $\vx$ but may be correlated between components
of $\vy = \boldsymbol{f}(\vx) + \epsilon$. The input variables are aggregated in the $d \times N$ design matrix $X$. After observing $Y$, the posterior mean $F_* \equiv \mathbb E(F(X_*))$ and covariance evaluated for validation data $X_{*}$ is given analytically by

\begin{align}
    F_{*} &= K(X_{*}, X) (K(X, X) + \Sigma_n)^{-1} Y, \\
    \label{eq:fpred}
    \textrm{cov}(F_{*}) &= 
        \begin{aligned}[t]
        & K(X_{*}, X_{*})  
         - K(X_{*}, X) (K(X, X) + \Sigma_n)^{-1} K(X, X_{*}),
    \end{aligned}
\end{align}
where $\Sigma_n \in \mathbb{R}^{D \times D}$ is the covariance matrix of the multivariate output noise. In the simplest case it is diagonal 
with entries $\Sigma^{ii}_n = \sigma_n^{\,2}$. Estimation of kernel parameters and $\Sigma_n$ 
given the input data is usually performed via optimization or sampling according 
to the marginal log-likelihood.

When a linear operator $\mathcal{L}$, e.g. differentiation, is applied to the Gaussian process, this yields a new Gaussian process \cite{eriksson2018, solak2003, raissi2017},
\begin{equation}
    \vg(\vx) = \mathcal{L} \vf(\vx) \sim \mathcal{GP}(\boldsymbol{l}(\vx), L(\vx,\vx^\prime)).
\end{equation}
Here the mean $\boldsymbol{l}(\vx)$ is given by $\boldsymbol{l}(\vx) = \mathcal{L} \vm(\vx)$ and a matrix-valued gradient kernel 
\begin{equation}
L(\vx, \vx^\prime) = (\mathcal{L}_{\vx} \otimes \mathcal{L}_{\vx^\prime})K(\vx, \vx^\prime) = \mathcal{L}_{\vx} K(\vx, \vx^\prime)\mathcal{L}^T_{\vx^\prime} 
\end{equation}
follows, where $\mathcal{L}^T_{\vx^\prime} $ is applied from the right to yield an outer product \cite{albert2020}. 

As differentiation is a linear operation, in particular the gradient of a Gaussian process over scalar functions $g(\vx)$ with kernel $k(\vx, \vx^\prime)$ remains a Gaussian process. The result is a multi-output GP where the covariance matrix is the Hessian of $K(\vx,\vx^\prime)$ containing all second derivatives in $(\vx,\vx^\prime)$. A joint GP, describing both, values and gradients is given by
\begin{equation}
    \begin{pmatrix}
      g(\vx) \\ {\nabla} g(\vx)
    \end{pmatrix} \sim \mathcal{GP}(\boldsymbol{n}(\vx), K(\vx, \vx^\prime)), 
\end{equation}
with $\boldsymbol{n}(\vx) = (m(\vx), \boldsymbol{l}(\vx))^T$ and where
\begin{equation}\label{eq:Ksym}
    K(\vx, \vx^\prime) = \begin{pmatrix}
      k(\vx, \vx^\prime) & k(\vx, \vx^\prime)\nabla^T_{\vx^\prime} \\
      \nabla_{\vx}k(\vx, \vx^\prime) & \nabla_{\vx} k(\vx, \vx^\prime)\nabla^T_{\vx^\prime} \\
          \end{pmatrix}
\end{equation}
contains $L(\vx, \vx^\prime)$ as the lower-right block. In the more general case of a linear operator $\mathcal{L}$, one may use the joint GP in Eq.~\ref{eq:Ksym} as a \textit{symmetric meshless formulation}~\cite{fasshauer1996} to find approximate solutions of the according linear (partial differential) equation.

\subsection{Symplectic GP regression}
To apply GP regression on symplectic maps we use Eq.~\ref{eq:Ksym} for the joint distribution of the generating function and its gradients in Eq.~\ref{eq:grad},

\begin{equation}
    \begin{pmatrix}\Tilde{F}(\vq,\vP) \\  \partial_{\vq} \Tilde{F}(\vq,\vP) \\   \partial_{\vP} \Tilde{F}(\vq,\vP) 
    \end{pmatrix} \sim \mathcal{GP}(\boldsymbol{n}(\vq,\vP), K(\vq,\vP,\vq^\prime,\vP^\prime))
\end{equation}
with
\begin{equation}
K(\vq,\vP,\vq^\prime,\vP^\prime) = 
    \begin{pmatrix}
        k & \partial_{\vq^\prime} k & \partial_{\vP^\prime} k\\
        \partial_{\vq} k & \partial_{\vq \vq^\prime} k & \partial_{\vq \vP^\prime} k\\
        \partial_{\vP}k & \partial_{\vP \vq^\prime} k & \partial_{\vP \vP^\prime} k 
    \end{pmatrix}.
\end{equation}
We cannot observe the generating function $\tilde{F}(\vq, \vP)$, but it is determined up to an additive constant via the predictor
\begin{equation}
    \Tilde{F}_{*} =  
    \begin{pmatrix}
     \partial_{\vq} k (X_*, X) \\
    \partial_{\vP} k (X_*, X)  \end{pmatrix} (L(X,X) + \Sigma_n)^{-1} Y,
    \label{eq:learnF}
\end{equation}
where column $i$ of $X$ for the $i$-th training orbit is composed of rows
\begin{equation}
    x_{1 \dots f,~i} = \vq_i\quad\text{and}\quad x_{(f+1) \dots 2f,~i} = \vP_i,
\end{equation} 
and similarly for $X_*$ and test points.  
Columns of $Y$ contain 
\begin{equation}
    y_{1 \dots f,~i} = \Delta \vq_i = \partial_{\vq} \Tilde{F}(\vq_i,\vP_i)\quad \text{and} \quad
    y_{(f+1) \dots 2f,~i} = -\Delta \vp_i = \partial_{\vP} \Tilde{F}(\vq_i,\vP_i).
\end{equation}
The matrix $L$ denotes the lower block 
\begin{equation}
L(\vq,\vP,\vq^\prime,\vP^\prime) = 
    \begin{pmatrix} 
        \partial_{\vq \vq^\prime} k & \partial_{\vq \vP^\prime} k\\
        \partial_{\vP \vq^\prime} k & \partial_{\vP \vP^\prime} k 
    \end{pmatrix}.
\end{equation}
This also allows to learn the Hamiltonian $H$ from Eq. \ref{eq:learnF} as for sufficiently small mapping times $H$ can be approximated by $\tilde{F}$ (up to a constant). 

For further investigations on temporal evolution of the Hamiltonian system and the construction of symplectic maps, we are interested in the gradients of $\Tilde{F}$ via the block $L$.
The predictive mean for this GP's output is given by
\begin{align}
    \begin{pmatrix}
    -\Delta \vp_* \\   
    \Delta \vq_*
    \end{pmatrix}
    =
    L(X_*, X) (L(X, X) + \Sigma_n)^{-1} 
    \begin{pmatrix}
        -\Delta \vp \\
        \Delta \vq
    \end{pmatrix}.
    \label{eq:mean}
\end{align}

Let's now check the symplecticity condition for predictors
$\vp_* = \vP_* - \Delta \vp_*(\vq_*, \vP_*)$ and 
$\vQ_* = \vq_* + \Delta \vq_*(\vq_*, \vP_*)$
according to Eq.~\ref{eq:sympl}. The derivatives of linear terms $\vP_*$ and $\vq_*$ vanish and by using Eq.~\ref{eq:mean}, the remaining derivatives
enter upper and lower rows of $L(X_*, X)$, respectively,

\begin{align}
    \frac{\partial \Delta\vq_*}{\partial \vq_*} - 
    \frac{\partial \Delta\vp_*}{\partial \vP_*} 
   &\propto
   \frac{\partial\begin{pmatrix}\partial_{\vP_* \vq}k & \partial_{\vP_* \vP}k\end{pmatrix}}
   {\partial \vq_*}
   - \frac{\partial\begin{pmatrix}\partial_{\vq_* \vq}k & \partial_{\vq_* \vP}k\end{pmatrix}}
   {\partial \vP_*}.
   \label{eq:zerosympl}
\end{align}
Due to symmetry of partial derivatives, the expected value of the symplecticity condition in Eq.~\ref{eq:zerosympl} is identically zero, so 
the predictive mean of Eq.~\ref{eq:mean} produces a symplectic map.
Due to the mixing of initial and final conditions by such a map, we can generally not
predict $\vQ_*$ and $\vP_*$ for a given $\vq_*, \vp_*$ right away.
Depending on the choice of the kernel, two cases have to be considered: 
\newline
\paragraph{Semi-implicit method}
In the most general case with a generating function $\Tilde{F}(\vq, \vP)$, equations for $\vP_*$
in Eq. \ref{eq:mean} are implicit and have to be solved iteratively as indicated in Algorithm \ref{al:implicit}. This corresponds to the implicit steps of a symplectic Euler integrator
in a non-separable Hamiltonian system \cite{Hairer2006}.

\begin{algorithm}
\SetAlgoLined
\textbf{Construction:}\\
Step 1: Usual GP regression of $\vP$ over initial $(\vq,\vp)$\\
Step 2: Symplectic GP regression of $-\Delta \vp$ and $\Delta \vq$ over mixed variables $(\vq,\vP)$ according to
    \begin{equation}
            \begin{pmatrix} -\Delta \vp \\ \Delta \vq 
    \end{pmatrix} \sim \mathcal{GP}(\boldsymbol{l}(\vq,\vP), L(\vq,\vP,\vq^\prime,\vP^\prime))
    \end{equation}
\newline
\textbf{Application:}\\
Step 3: Initial guess $\vP_*(\vq_*, \vp_*)$ from GP of Step 1 \\
Step 4: Solve implicit equation in $\vP_*$ via 

\begin{equation}
    \Delta \vp_*(\vq_*, \vP_*) - (\vp_* - \vP_*) = 0,
\end{equation}
    
predicting $\Delta \vp_*$ via Eq. \ref{eq:mean} from symplectic GP of Step 2.\\

Step 4: Explicitly evaluate 
\begin{equation}
    \vQ_* = \vq_* + \Delta \vq_*(\vq_*, \vP_*),
\end{equation}
predicting $\Delta \vq_*$ via Eq.~\ref{eq:mean} from symplectic GP of Step 2.\\
\caption{Semi-implicit symplectic GP map}
\label{al:implicit}
\end{algorithm}
\newpage
\paragraph{Explicit method}

When considering a generating function in separable form $\Tilde{F}(\vq, \vP) = V(\vq) + T(\vP)$, the resulting transformation equations reduce to 

\begin{eqnarray}
    \vp (\vq, \vP) = \frac{\partial \Tilde{F}(\vq, \vP)}{\partial \vq} = \frac{\partial V(\vq)}{\partial \vq}, \\
    \vQ (\vq, \vP) = \frac{\partial \Tilde{F}(\vq, \vP)}{\partial \vP} = \frac{\partial T(\vP)}{\partial \vP}, 
\end{eqnarray}

resulting in a simplified lower block $L(\vq, \vP, \vq^\prime, \vP^\prime)$ with off-diagonal elements equal to 0:

\begin{equation}
L(\vq,\vP,\vq^\prime,\vP^\prime) = 
    \begin{pmatrix} 
        \partial_{\vq \vq^\prime} k & 0\\
        0 & \partial_{\vP \vP^\prime} k 
    \end{pmatrix}.
\end{equation}

This choice of generating function results in a simplified construction and application of the symplectic map as explained in Algorithm \ref{al:explicit}. This corresponds to the case of a 
separable Hamiltonian where the symplectic Euler method becomes fully explicit. The hope is that
such a separable approximation is still possible also for non-separable systems, thereby resulting
in a trade-off between performance and accuracy compared to the semi-implicit variant.

\begin{algorithm}
\SetAlgoLined
\textbf{Construction:}\\
Step 1: GP regression of $-\Delta \vp$ and $\Delta \vq$ over mixed variables $(\vq,\vP)$ according to
    \begin{equation}
            \begin{pmatrix} -\Delta \vp \\ \Delta \vq 
    \end{pmatrix} \sim \mathcal{GP}(\boldsymbol{l}(\vq,\vP), L(\vq,\vP,\vq^\prime,\vP^\prime))
    \end{equation}
\newline
\textbf{Application:}\\
Step 2: Solve explicit equation in $\vP_*$ via 

\begin{equation}
    \Delta \vp_*(\vq_*) - (\vp_* - \vP_*) = 0,
\end{equation}
    
with $\Delta \vp_*(\vq_*)$ predicted from GP of Step 1.\\
    
Step 3: Evaluate 
\begin{equation}
    \vQ_* = \vq_* + \Delta \vq_*(\vP_*),
\end{equation}
with $\Delta \vq_*(\vP_*)$ predicted from GP of Step 1.\\
\caption{Explicit symplectic GP map}
\label{al:explicit}
\end{algorithm}

\section{Numerical experiments}
The described symplectic regression methods for flow maps are benchmarked for two 
Hamiltonian systems: the pendulum and the Hénon-Heiles system. In addition, also the Hamiltonian function $H$ is predicted using GPs following Eq. \ref{eq:learnF}. 
We compare
\begin{enumerate}
    \item implicit symplectic GP (Algorithm \ref{al:implicit}),
    \item explicit symplectic GP (Algorithm \ref{al:explicit}),
    \item implicit symmetric regression with a spectral basis,
    \item implicit non-symmetric collocation with RBFs.
\end{enumerate}

To assess the quality and stability of the proposed mapping methods, two quality measures are used. 
The geometric distance is computed to compare the first application of the constructed map step to the respective timestep of a reference orbit in phase space. This phase space distance is given by
\begin{equation}
    g = \sqrt{(\vq - \vq_{\mathrm{ref}})^2 + (\vp - \vp_{\mathrm{ref}})^2}, 
    \label{eq:dist}
\end{equation} 
where $\vq_{\mathrm{ref}}$ and $\vp_{\mathrm{ref}}$ denote the reference orbits and $\vq$ and $\vp$ the mapped orbits. The reference orbits are calculated using an adaptive step-size Runge-Kutta scheme. 

As energy is a constant of motion and should be preserved, the normalized energy 
oscillation given by
\begin{equation}
    E_{\mathrm{osc}} = \frac{\textrm{Std}(\Bar{H})}{\Bar{H}}, 
    \label{eq:Eosc}
\end{equation}
where $\Bar{H}$ is the mean, serves as a criterion for mapping quality. Here, the energy oscillations are averaged over the whole considered time: $k = 300$ subsequent applications 
of the map. 

\subsection{Pendulum}

\begin{figure}
\begin{center}
\includegraphics[width=0.65\linewidth]{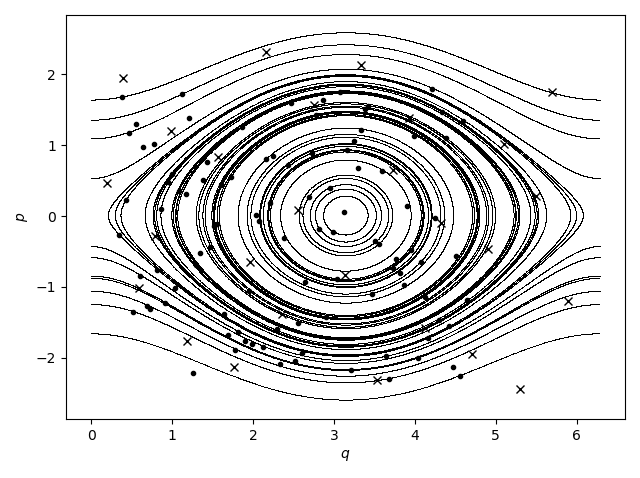}
\caption{Reference orbits of pendulum in phase space for 50 initial conditions (validation data, denoted with dots) with the position of 27 training data points (denoted with x) sampled by a Halton sequence.}
\label{fig:phasespace}
\end{center}
\end{figure}

The pendulum is a nonlinear oscillator with position $q$ and momentum $p$ and exhibits two kinds of motion: libration and rotation, which are separated by the separatrix in phase space (Fig. \ref{fig:phasespace}). The system of a pendulum corresponds to a particle trapped in a cosine potential \cite{lichtenberg1992}. The Hamiltonian is given by 

\begin{equation}
    H(q,p) = \frac{p^2}{2} + U_0 ( 1 - \textrm{cos}(q)),
    \label{eq:hamiltonian}
\end{equation}
from which the equations of motion follow,

\begin{eqnarray}
    \dot{q} &=& \frac{\partial H}{\partial p} = p, \\
    \dot{p} &=& - \frac{\partial H}{\partial q} = - \textrm{sin}(q).
\end{eqnarray}

The underlying periodic topology suggest the choice of a periodic kernel function in $q$ and a squared exponential kernel function in $p$:

\begin{align}
    V(q, q_i) &\propto  \textrm{exp} \left( -\frac{\textrm{sin}^2((q-q_i)/2)}{2 l_q^2} \right),\\
    T(P, P_i) &\propto \textrm{exp} \left( -\frac{(P-P_i)^2}{2 l_p^2} \right).
\end{align}
The hyperparameters, $l_q$ and $l_p$ that correspond to the length scales in $q$ and $p$ respectively, are set to optimized maximum likelihood values by minimizing the negative log-likelihood \cite{rasmussen2005}. For the product kernel $k(q,q_i, P, P_i) = \sigma_f^2 V(q, q_i) T(P, P_i)$ used in the implicit method, the noise in observations $\sigma_n^2$ (Eq. \ref{eq:fpred}) is set to $10^{-16}$, whereas for the sum kernel $k(q,q_i, P, P_i) = \sigma_f^2 (V(q, q_i) +T(P, P_i))$ for the explicit method, $\sigma_n^2 = 10^{-10}$. The scaling of the kernel $\sigma_f^2$ that quantifies the amplitude of the fit is set in accordance with the observations is fixed at $2\,\mathrm{max}(|Y|)^2$, where $Y$ corresponds to the training output. 
For the spectral method, an expansion in Fourier sine/cosine harmonics in $q$ and Hermite polynomials in $p$ was chosen to account for periodic phase space. The number of modes and the degree depend on the number of training points to obtain a fully or overdetermined fit.  

\begin{figure}[ht]
      \includegraphics[width=0.8\linewidth]{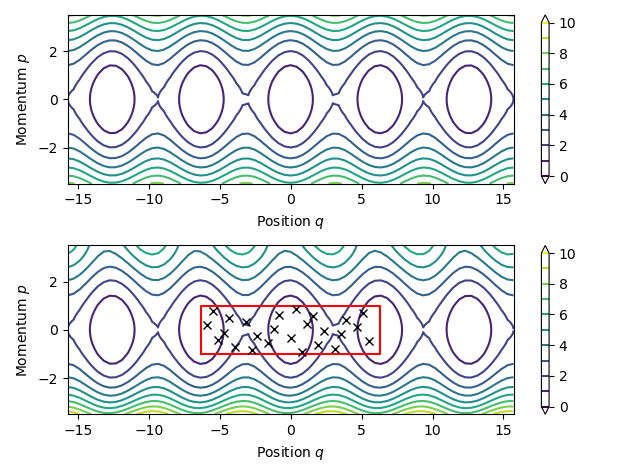}
      
\caption{Upper: Hamiltonian function of the pendulum (Eq. \ref{eq:hamiltonian}), lower: Approximation using symplectic Gaussian process regression of the generating function via Eq.~\ref{eq:learnF}. The 25 training points are denoted by $\times$. Excellent interpolation
performance and surprisingly good extrapolation are observed.}
\label{fig:learnedH}
\end{figure}

The Hamiltonian function $H$ given in Eq. \ref{eq:hamiltonian} is learned from the initial $(q,p)$ and final state $(Q,P)$ using Eq. \ref{eq:learnF}. In contrast to earlier proposed methods \cite{bertalan2019}, derivatives of $H$ are not needed explicitly as the training data consists only of observable states of the dynamical system in time. We use a fully symmetric collocation scheme with a product kernel $k(q,q_i, P, P_i) \propto V(q, q_i) T(P, P_i)$. In Fig. \ref{fig:learnedH}, the Hamiltonian function calculated exactly from Eq. \ref{eq:hamiltonian} is compared to the approximation using the implicit symplectic GP method trained with 25 training points sampled from a Halton sequence within the range $q \in [-2 \pi, 2 \pi]$ and $p \in [-1.0, 1.0]$. The approximation is validated using 5625 random points within the same range. Using the mean squared error to evaluate the losses, we get for the training loss $1.3 \cdot 10^{-5}$ and for the validation loss $6.3 \cdot 10^{-5}$. As evident in Fig. \ref{fig:learnedH}, the extrapolation ability is restricted to areas close to the range of the training data in $p$, whereas longer extrapolations in $q$ are possible due to the periodicity of the system and the use of a periodic kernel function. 

To evaluate the different mapping methods, we use training data points sampled from a Halton sequence within the range $q \in [0, 2 \pi]$ and $p \in [-2.5, 2.5]$. 
The 100 validation data points as shown in Fig. \ref{fig:phasespace} are chosen randomly within the range $q \in [\pi - 2.8, \pi + 1.5]$ and $p \in [-2.3, 1.8]$ and include several possible unstable points as they are close to the separatrix. Naturally, motion near the separatrix is unstable, as a small deviation from the true orbit can result in a change of kind of motion. As this does not represent physical behaviour, those unstable points are removed from the final results. 

\subsection{Hénon-Heiles Potential}
The Hénon-Heiles problem is a classical example of nonlinear Hamiltonian systems with $f=2$ degrees of freedom and a 4D phase space~\cite{lichtenberg1992}. The Hamiltonian for the Hénon-Heiles problem is given by
\begin{equation}
    H(\vq, \vp) = \frac{1}{2}(q_1^2 + q_2^2) + \frac{1}{2}(p_1^2 + p_2^2) + \lambda (q_1^2 q_2 - \frac{1}{3}q_2^3),
    \label{eq:henonHam}
\end{equation}
where we will fix $\lambda = 1$. The equations of motion follow directly as
\begin{eqnarray}
    \dot{q_1} &=& \frac{\partial H}{\partial p_1} = p_1, \\
    \dot{q_2} &=& \frac{\partial H}{\partial p_2} = p_2, \\
    \dot{p_1} &=& - \frac{\partial H}{\partial q_1} = - q_1 - 2 \lambda q_1 q_2, \\
    \dot{p_2} &=& - \frac{\partial H}{\partial q_2} = - q_2 - \lambda (q_1^2 - q_2^2).
\end{eqnarray}
The underlying potential continuously varies from a harmonic potential for small values of $q_1$ and $q_2$ to triangular equipotential lines on the edges. For energies lower than the limiting potential energy $H_{\mathrm{esc}} =  1/6$, the orbit is trapped within the potential. However for larger energies, three escape channels appear due to the special shape of the potential, through which the orbit may escape \cite{Zotos}. Therefore, the training and validation data is set to a restricted area in phase space in order to keep the motion bounded within the potential. 

Here, a squared exponential kernel function is used in all dimensions, where the hyperparameter $l$ is set to its optimized maximium likelihood value. The noise in observations $\sigma_n^2$ (Eq. \ref{eq:fpred}) is set to $10^{-16}$ for the implicit method and to $10^{-10}$ for the explicit method. As in the pendulum case, $\sigma_f^2$ is set in accordance with the observations to $2\,\textrm{max}(|Y|)^2$, where $Y$ corresponds to the change in coordinates. 
For the spectral method, an expansion Hermite polynomials in all dimensions was chosen to represent the non-periodic structure of phase space. The degree depends on the number of training points to obtain a fully or overdetermined fit. 

\begin{figure}
      \includegraphics[width=0.8\linewidth]{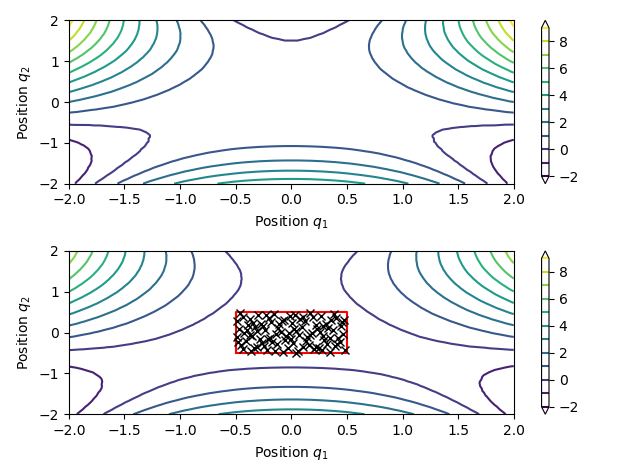}
      
\caption{Upper: Hamiltonian function of the Hénon-Heiles system (Eq. \ref{eq:henonHam}), lower: Approximation using symplectic Gaussian process regression of the generating function via Eq.~\ref{eq:learnF}. The projection of the 50 training points onto the plane $\vp = 0$ is denoted by x. As in the pendulum case, the Hamiltonian function $H$ is also extrapolated
well relatively far from the training region.}
\label{fig:learnedHhenon}
\end{figure}

The Hamiltonian function (Eq. \ref{eq:henonHam}) is learned from 250 training data points sampled from a Halton sequence in the range $\vq \in [-0.5, 0.5]$ and $\vp \in [-0.5, 0.5]$ for the initial $(\vq,\vp)$ and final state $(\vQ,\vP)$ using Eq. \ref{eq:learnF}. In Fig. \ref{fig:learnedHhenon}, the Hamiltonian function calculated from Eq. \ref{eq:henonHam} is compared to the approximation using GPs. The approximation is validated using 46656 points within the same range as the training points. Using the mean squared error to evaluate the losses, we get for the training loss $6.9 \cdot 10^{-5}$ and for the validation loss $8.1 \cdot 10^{-5}.$

For the application of the symplectic mapping methods, training data points are sampled from a Halton sequence in the range $\vq \in [-0.5, 0.5]$ and $\vp \in [-0.5, 0.5]$. The 100 validation data points are chosen randomly within the range $\vq \in [-0.2, 0.2]$ and $\vp \in [-0.2, 0.2]$.

\section{Discussion}

\begin{figure}
\begin{minipage}[t]{0.49\linewidth}
     \includegraphics[width=\linewidth]{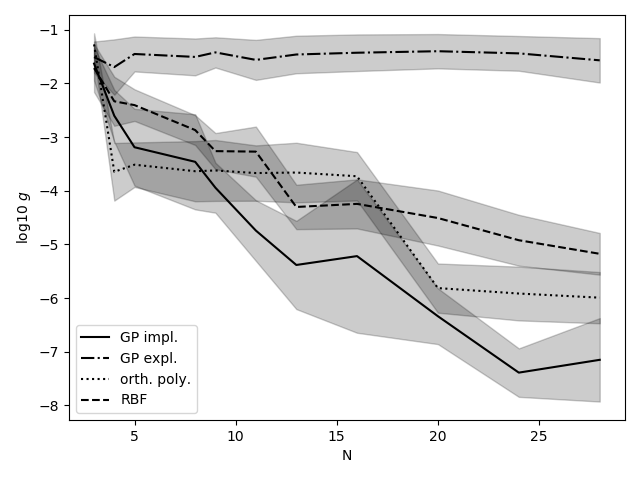}
\end{minipage}
\begin{minipage}[t]{0.49\linewidth}
    \includegraphics[width=\linewidth]{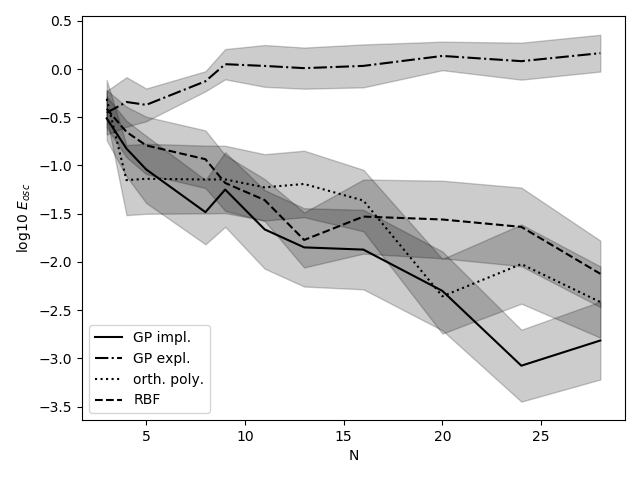}
\end{minipage}
\caption{Pendulum: Comparison of geometrical distance (Eq. \ref{eq:dist}) (left) and normalized energy oscillations (Eq. \ref{eq:Eosc}) (right) of implicit and explicit symplectic methods with implicit and explicit GP, RBF and Fourier-Hermite basis functions for variable number of training points $N$, sampled using a Halton sequence, and a fixed mapping time $t = 0.5 \tau_b$, where $\tau_b$ is the linear bounce time. The grey areas surrounding the mean correspond to the standard deviation for 100 validation points.}
\label{fig:constt}
\end{figure}

\begin{figure}[ht]
\begin{minipage}[t]{0.49\linewidth}
     \includegraphics[width=\linewidth]{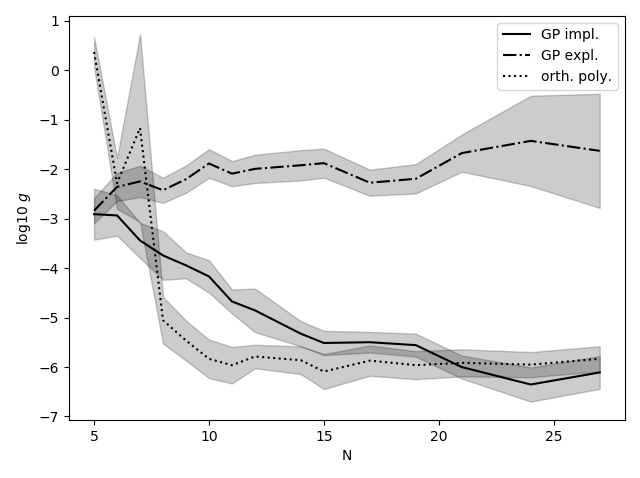}   
\end{minipage}
\begin{minipage}[t]{0.49\linewidth}
    \includegraphics[width=\linewidth]{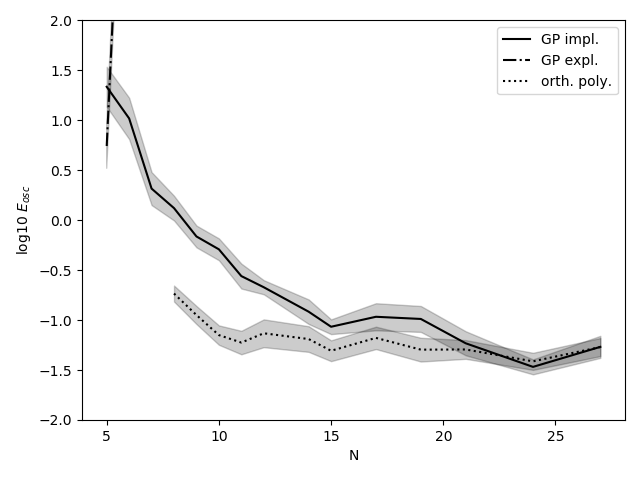}
\end{minipage}
\caption{Hénon-Heiles system: Comparison of geometrical distance (Eq. \ref{eq:dist}) (left)  normalized energy oscillations (Eq. \ref{eq:Eosc}) (right) of implicit and explicit symplectic methods with GPs and Hermite basis functions for variable number of training points $N$, sampled using a Halton sequence, and a fixed mapping time $t = 0.5 \tau_b$. The results for the explicit method with GPs are cut as the method becomes unstable for larger mapping times.}
\label{fig:henonNm}
\end{figure}

In Fig. \ref{fig:constt} and Fig. \ref{fig:henonNm}, the four methods are compared for the one dimensional pendulum and the Hénon-Heiles problem, respectively, using the quality criteria given in Eq. \ref{eq:dist} and Eq. \ref{eq:Eosc} for a fixed mapping time $t/\tau_b$, where $\tau_b$ is the linearized bounce time, but a variable number of training points $N$. As expected, the geometric distance $g$ as well as the energy oscillation $E_{\mathrm{osc}}$ are increasing for a increasing mapping time. Since no Newton iterations are needed, the explicit GP method is faster than the implicit method in it's region of validity. As for the first guess for Newton's method a separate GP is used as indicated in Algorithm \ref{al:implicit}, less then 5 Newton iterations are typically necessary in the implicit case. 

\begin{figure}
      \includegraphics[width=0.6\linewidth]{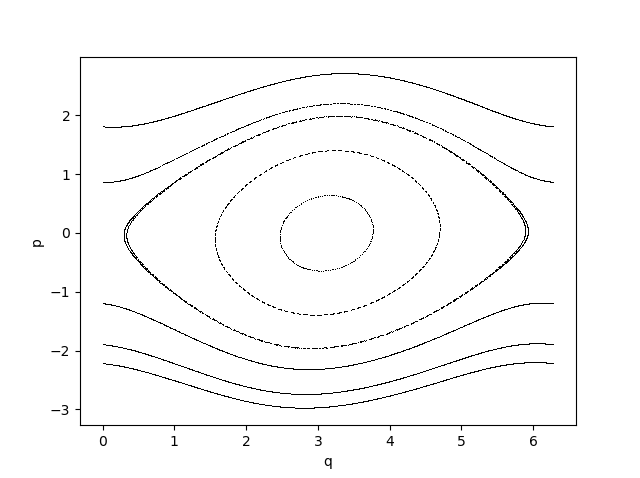}
      
\caption{Orbits in phase space using the explicit method with GPs. $N = 20$, $t = 0.08\tau_b$.}
\label{fig:explEuler}
\end{figure}

As apparent in Fig.~\ref{fig:explEuler} for the 1D case, the orbits in phase space resulting from the explicit method with GPs are tilted which explains the bad performance regarding the geometrical distance. For smaller mapping times, the tilt and therefore also the energy oscillation reduces. This is in accordance with similar behavior of explicit-implicit Euler integration schemes. More severely, the explicit GP loses long-term stability at increasing mapping time $t/\tau_b$ in the Hénon-Heiles system due to certain orbits that escape the trapped state after a few 10-100 applications of the map. This severely limits the
applicability range of the explicit GP method in its current state.

Spectral linear regression produces very accurate results for very small mapping times, as the interpolated data (the change in coordinates) is almost 0, and the generating function inherits the polynomial structure of the Hamiltonian $H$ that can be fitted exactly. At larger mapping times, implicit GP regression and spectral methods perform similarly, with somewhat better results of the implicit GP for the pendulum, and higher accuracy of the spectral fit
for the Hénon-Heiles system.

\begin{figure}
\begin{minipage}[t]{0.49\linewidth}
     \includegraphics[width=\linewidth]{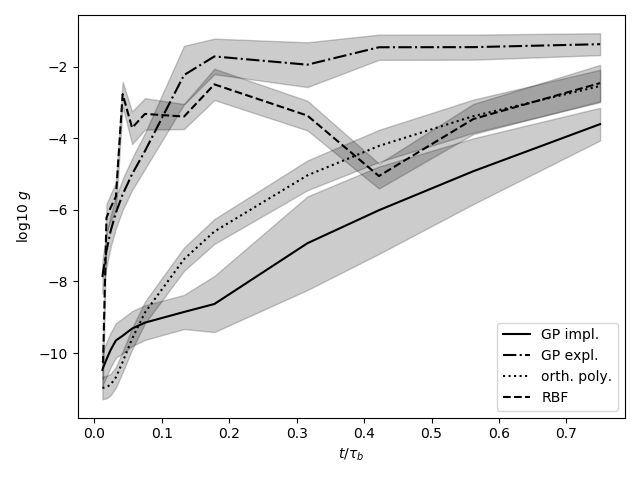}
\end{minipage}
\begin{minipage}[t]{0.49\linewidth}
    \includegraphics[width=\linewidth]{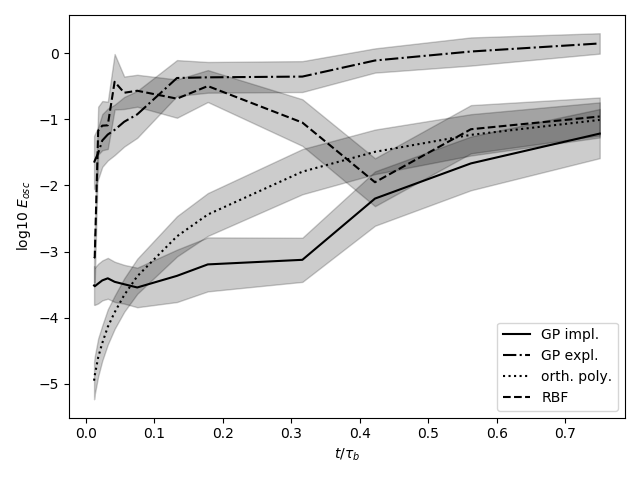}
\end{minipage}
\caption{Pendulum: Comparison of geometrical distance (Eq. \ref{eq:dist}) (left) and normalized energy oscillations (Eq. \ref{eq:Eosc}) (right) of implicit and explicit symplectic methods with implicit and explicit GP, RBF and Fourier-Hermite basis functions for a fixed number of training points $N = 15$, sampled using a Halton sequence, and a variable mapping time $t/\tau_b$, where $\tau_b$ is the linear bounce time. The grey areas surrounding the mean correspond to the standard deviation for 100 validation points.}
\label{fig:constN}
\end{figure}

\begin{figure}
\begin{minipage}[t]{0.49\linewidth}
     \includegraphics[width=\linewidth]{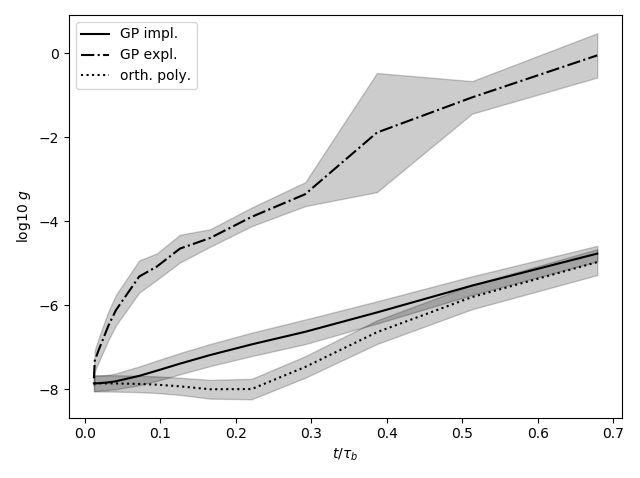}
\end{minipage}
\begin{minipage}[t]{0.49\linewidth}
    \includegraphics[width=\linewidth]{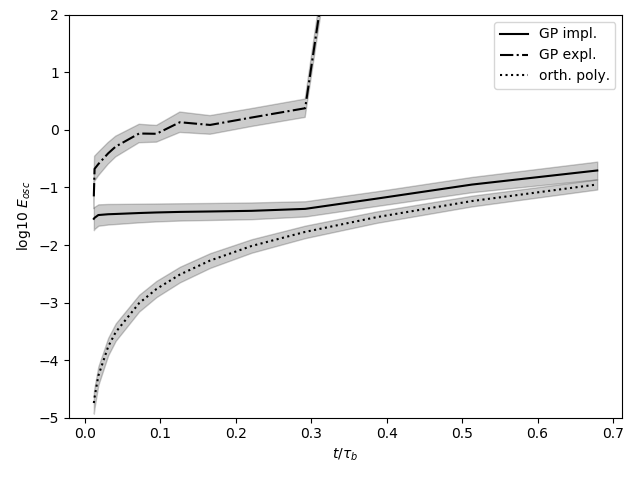}
\end{minipage}
\caption{Hénon-Heiles system: Comparison of geometrical distance (Eq. \ref{eq:dist}) (left) and normalized energy oscillations (Eq. \ref{eq:Eosc}) (right) of implicit and explicit symplectic methods with GPs and Hermite basis functions for a fixed number of training points $N = 20$, sampled using a Halton sequence, and a variable mapping time $t/\tau_b$. The grey areas surrounding the mean correspond to the standard deviation for 100 validation points. The results for the explicit method with GPs are cut as the method becomes unstable for larger mapping times. }
\label{fig:henonN}
\end{figure}

To investigate the behavior with increasing number of $N$, in Fig. \ref{fig:constN} for the pendulum and Fig. \ref{fig:henonN} for the Hénon-Heiles system, the quality measures are compared for fixed mapping time $t/\tau_b$, but with a variable number of training points. The explicit method does not improve with $N$ due to the tilt in phase space, that is an inherent structural feature of the forced splitting into a sum kernel that cannot be fixed by adding more training data. The implicit methods improve considerably with $N$. While the explicit
GP method remains stable in the pendulum, it again becomes unstable already at a small
number $N$ below $10$ for the Hénon-Heiles system. The visible steps for the implicit method with spectral basis arises from the used number of modes that depends on $N$. 

\section{Conclusion and Outlook}

In this paper, we have presented a novel approach to represent symplectic flow maps of Hamiltonian dynamical systems using Gaussian Process regression. Intended applications are
long-term tracing of fast charged particles in accelerators and magnetic plasma confinement
configurations.
A considerable advantage compared to existing methods in a spline or spectral basis is the possibility of using input data of arbitrary geometry with GPs. Moreover, the presented method uses considerably less training data than neural networks in the presented test cases. 
The concept was validated on two Hamiltonian systems: the 1D pendulum and the 2D Hénon-Heiles problem. An implicit  approach was shown to yield similar accuracy to linear regression 
in a spectral basis, whereas an explicit mapping requires no iterations in application 
of the map at the cost of accuracy and stability. Observation of training data within a short period of time allows
for an accurate interpolation and even extrapolation of the Hamiltonian function $H$
using substantially less training points compared to existing methods.

To increase the accuracy of the symplectic mappings as well as the prediction of $H$, higher order implicit methods analogous to symplectic schemes such as as midpoint, Gauss-Legendre or higher order RK schemes, could be investigated in the future. Especially the explicit method in combination with a Verlet scheme seems promising to leverage fast computation and the possible higher accuracy. This may also aid to overcome stability issues faced in the Hénon-Heiles system due to reduced training error with less risk of overfitting. Another important
next step could be the investigation of maps between Poincaré sections. This will allow to 
increase mapping time even further, but require prior separation of regions with different
orbit classes in phase space. In the future higher dimensional systems as well as chaotic behavior could be investigated based on such Poincaré maps.

\section{Acknowledgements}
The authors would like to thank Johanna Ganglbauer and Bob Warnock for insightful discussions
on spline methods for interpolating symplectic maps and Manal Khallaayoune for supporting implementation tests.
The present contribution is supported by the Helmholtz Association of German Research Centers under the joint research school HIDSS-0006 "Munich School for Data Science - MUDS" and the Reduced Complexity grant No. ZT-I-0010.

\nocite{*}
\bibliography{SGPR}

\providecommand{\noopsort}[1]{}\providecommand{\singleletter}[1]{#1}%
\begin{thebibliography}{29}%
\makeatletter
\providecommand \@ifxundefined [1]{%
 \@ifx{#1\undefined}
}%
\providecommand \@ifnum [1]{%
 \ifnum #1\expandafter \@firstoftwo
 \else \expandafter \@secondoftwo
 \fi
}%
\providecommand \@ifx [1]{%
 \ifx #1\expandafter \@firstoftwo
 \else \expandafter \@secondoftwo
 \fi
}%
\providecommand \natexlab [1]{#1}%
\providecommand \enquote  [1]{``#1''}%
\providecommand \bibnamefont  [1]{#1}%
\providecommand \bibfnamefont [1]{#1}%
\providecommand \citenamefont [1]{#1}%
\providecommand \href@noop [0]{\@secondoftwo}%
\providecommand \href [0]{\begingroup \@sanitize@url \@href}%
\providecommand \@href[1]{\@@startlink{#1}\@@href}%
\providecommand \@@href[1]{\endgroup#1\@@endlink}%
\providecommand \@sanitize@url [0]{\catcode `\\12\catcode `\$12\catcode
  `\&12\catcode `\#12\catcode `\^12\catcode `\_12\catcode `\%12\relax}%
\providecommand \@@startlink[1]{}%
\providecommand \@@endlink[0]{}%
\providecommand \url  [0]{\begingroup\@sanitize@url \@url }%
\providecommand \@url [1]{\endgroup\@href {#1}{\urlprefix }}%
\providecommand \urlprefix  [0]{URL }%
\providecommand \Eprint [0]{\href }%
\providecommand \doibase [0]{http://dx.doi.org/}%
\providecommand \selectlanguage [0]{\@gobble}%
\providecommand \bibinfo  [0]{\@secondoftwo}%
\providecommand \bibfield  [0]{\@secondoftwo}%
\providecommand \translation [1]{[#1]}%
\providecommand \BibitemOpen [0]{}%
\providecommand \bibitemStop [0]{}%
\providecommand \bibitemNoStop [0]{.\EOS\space}%
\providecommand \EOS [0]{\spacefactor3000\relax}%
\providecommand \BibitemShut  [1]{\csname bibitem#1\endcsname}%
\let\auto@bib@innerbib\@empty
\bibitem [{\citenamefont {Goldstein}(1980)}]{Goldstein1980}%
  \BibitemOpen
  \bibfield  {author} {\bibinfo {author} {\bibfnamefont {H.}~\bibnamefont
  {Goldstein}},\ }\href@noop {} {\emph {\bibinfo {title} {Classical
  Mechanics}}},\ \bibinfo {edition} {2nd}\ ed.\ (\bibinfo  {publisher}
  {Addison-Wesley},\ \bibinfo {year} {1980})\BibitemShut {NoStop}%
\bibitem [{\citenamefont {Arnold}(1989)}]{Arnold1997}%
  \BibitemOpen
  \bibfield  {author} {\bibinfo {author} {\bibfnamefont {V.~I.}\ \bibnamefont
  {Arnold}},\ }\href {http://link.springer.com/10.1007/978-1-4757-2063-1}
  {\emph {\bibinfo {title} {{Mathematical Methods of Classical Mechanics}}}},\
  \bibinfo {series} {Graduate Texts in Mathematics}, Vol.~\bibinfo {volume}
  {60}\ (\bibinfo  {publisher} {Springer},\ \bibinfo {address} {New York, NY},\
  \bibinfo {year} {1989})\BibitemShut {NoStop}%
\bibitem [{\citenamefont {Marsden}\ and\ \citenamefont
  {Ratiu}(1999)}]{Marsden1999}%
  \BibitemOpen
  \bibfield  {author} {\bibinfo {author} {\bibfnamefont {J.~E.}\ \bibnamefont
  {Marsden}}\ and\ \bibinfo {author} {\bibfnamefont {T.~S.}\ \bibnamefont
  {Ratiu}},\ }\href@noop {} {\emph {\bibinfo {title} {Introduction to Mechanics
  and Symmetry}}}\ (\bibinfo  {publisher} {Springer},\ \bibinfo {address} {New
  York},\ \bibinfo {year} {1999})\BibitemShut {NoStop}%
\bibitem [{\citenamefont {Neil}(2011)}]{Neil2011}%
  \BibitemOpen
  \bibfield  {author} {\bibinfo {author} {\bibfnamefont {R.~M.}\ \bibnamefont
  {Neil}},\ }\bibfield  {title} {\enquote {\bibinfo {title} {{MCMC} using
  {Hamiltonian} dynamics},}\ }in\ \href@noop {} {\emph {\bibinfo {booktitle}
  {Handbook of Markov Chain Monte Carlo}}},\ \bibinfo {editor} {edited by\
  \bibinfo {editor} {\bibfnamefont {S.}~\bibnamefont {Brooks}}, \bibinfo
  {editor} {\bibfnamefont {A.}~\bibnamefont {Gelman}}, \bibinfo {editor}
  {\bibfnamefont {G.~L.}\ \bibnamefont {Jones}}, \ and\ \bibinfo {editor}
  {\bibfnamefont {X.-L.}\ \bibnamefont {Meng}}}\ (\bibinfo  {publisher}
  {Chapman {\&} Hall/CRC},\ \bibinfo {year} {2011})\ pp.\ \bibinfo {pages}
  {113--162}\BibitemShut {NoStop}%
\bibitem [{\citenamefont {Lichtenberg}\ and\ \citenamefont
  {Lieberman}(1992)}]{lichtenberg1992}%
  \BibitemOpen
  \bibfield  {author} {\bibinfo {author} {\bibfnamefont {A.}~\bibnamefont
  {Lichtenberg}}\ and\ \bibinfo {author} {\bibfnamefont {M.}~\bibnamefont
  {Lieberman}},\ }\href {https://books.google.de/books?id=2ssPAQAAMAAJ} {\emph
  {\bibinfo {title} {Regular and chaotic dynamics}}},\ Applied mathematical
  sciences\ (\bibinfo  {publisher} {Springer-Verlag},\ \bibinfo {year}
  {1992})\BibitemShut {NoStop}%
\bibitem [{\citenamefont {Hairer}, \citenamefont {Lubich},\ and\ \citenamefont
  {Wanner}(2006)}]{Hairer2006}%
  \BibitemOpen
  \bibfield  {author} {\bibinfo {author} {\bibfnamefont {E.}~\bibnamefont
  {Hairer}}, \bibinfo {author} {\bibfnamefont {C.}~\bibnamefont {Lubich}}, \
  and\ \bibinfo {author} {\bibfnamefont {G.}~\bibnamefont {Wanner}},\ }\href
  {https://link.springer.com/book/10.1007/3-540-30666-8} {\emph {\bibinfo
  {title} {{Geometric numerical integration: structure-preserving algorithms
  for ordinary differential equations}}}}\ (\bibinfo  {publisher} {Springer},\
  \bibinfo {year} {2006})\BibitemShut {NoStop}%
\bibitem [{\citenamefont {Abdullaev}(2006)}]{Abdullaev2006}%
  \BibitemOpen
  \bibfield  {author} {\bibinfo {author} {\bibfnamefont {S.~S.}\ \bibnamefont
  {Abdullaev}},\ }\href@noop {} {\emph {\bibinfo {title} {Construction of
  Mappings for {Hamiltonian} Systems and Their Applications}}}\ (\bibinfo
  {publisher} {Springer},\ \bibinfo {year} {2006})\BibitemShut {NoStop}%
\bibitem [{\citenamefont {Berg}\ \emph {et~al.}(1994)\citenamefont {Berg},
  \citenamefont {Warnock}, \citenamefont {Ruth},\ and\ \citenamefont
  {Forest}}]{Berg1994}%
  \BibitemOpen
  \bibfield  {author} {\bibinfo {author} {\bibfnamefont {J.~S.}\ \bibnamefont
  {Berg}}, \bibinfo {author} {\bibfnamefont {R.~L.}\ \bibnamefont {Warnock}},
  \bibinfo {author} {\bibfnamefont {R.~D.}\ \bibnamefont {Ruth}}, \ and\
  \bibinfo {author} {\bibfnamefont {{\'{E}}.}~\bibnamefont {Forest}},\
  }\bibfield  {title} {\enquote {\bibinfo {title} {Construction of symplectic
  maps for nonlinear motion of particles in accelerators},}\ }\href {\doibase
  10.1103/PhysRevE.49.722} {\bibfield  {journal} {\bibinfo  {journal} {Physical
  Review E}\ }\textbf {\bibinfo {volume} {49}},\ \bibinfo {pages} {722--739}
  (\bibinfo {year} {1994})}\BibitemShut {NoStop}%
\bibitem [{\citenamefont {Kasilov}, \citenamefont {Moiseenko},\ and\
  \citenamefont {Heyn}(1997)}]{Kasilov1997}%
  \BibitemOpen
  \bibfield  {author} {\bibinfo {author} {\bibfnamefont {S.~V.}\ \bibnamefont
  {Kasilov}}, \bibinfo {author} {\bibfnamefont {V.~E.}\ \bibnamefont
  {Moiseenko}}, \ and\ \bibinfo {author} {\bibfnamefont {M.~F.}\ \bibnamefont
  {Heyn}},\ }\bibfield  {title} {\enquote {\bibinfo {title} {Solution of the
  drift kinetic equation in the regime of weak collisions by stochastic mapping
  techniques},}\ }\href {\doibase 10.1063/1.872223} {\bibfield  {journal}
  {\bibinfo  {journal} {Phys. Plasmas}\ }\textbf {\bibinfo {volume} {4}},\
  \bibinfo {pages} {2422} (\bibinfo {year} {1997})}\BibitemShut {NoStop}%
\bibitem [{\citenamefont {Kasilov}\ \emph {et~al.}(2002)\citenamefont
  {Kasilov}, \citenamefont {Kernbichler}, \citenamefont {Nemov},\ and\
  \citenamefont {Heyn}}]{Kasilov2002}%
  \BibitemOpen
  \bibfield  {author} {\bibinfo {author} {\bibfnamefont {S.~V.}\ \bibnamefont
  {Kasilov}}, \bibinfo {author} {\bibfnamefont {W.}~\bibnamefont
  {Kernbichler}}, \bibinfo {author} {\bibfnamefont {V.~V.}\ \bibnamefont
  {Nemov}}, \ and\ \bibinfo {author} {\bibfnamefont {M.~F.}\ \bibnamefont
  {Heyn}},\ }\bibfield  {title} {\enquote {\bibinfo {title} {Mapping technique
  for stellarators},}\ }\href {\doibase 10.1063/1.1493793} {\bibfield
  {journal} {\bibinfo  {journal} {Phys. Plasmas}\ }\textbf {\bibinfo {volume}
  {9}},\ \bibinfo {pages} {3508} (\bibinfo {year} {2002})}\BibitemShut
  {NoStop}%
\bibitem [{\citenamefont {Warnock}\ \emph {et~al.}(2009)\citenamefont
  {Warnock}, \citenamefont {Cai}, \citenamefont {Ellison},\ and\ \citenamefont
  {Mexico}}]{warnock2009}%
  \BibitemOpen
  \bibfield  {author} {\bibinfo {author} {\bibfnamefont {R.}~\bibnamefont
  {Warnock}}, \bibinfo {author} {\bibfnamefont {Y.}~\bibnamefont {Cai}},
  \bibinfo {author} {\bibfnamefont {J.~A.}\ \bibnamefont {Ellison}}, \ and\
  \bibinfo {author} {\bibfnamefont {N.}~\bibnamefont {Mexico}},\ }\bibfield
  {title} {\enquote {\bibinfo {title} {Construction of large period symplectic
  maps by interpolative methods},}\ \ }(\bibinfo {year} {2009})\BibitemShut
  {NoStop}%
\bibitem [{\citenamefont {Bertalan}\ \emph {et~al.}(2019)\citenamefont
  {Bertalan}, \citenamefont {Dietrich}, \citenamefont {Mezi{\'{c}}},\ and\
  \citenamefont {Kevrekidis}}]{bertalan2019}%
  \BibitemOpen
  \bibfield  {author} {\bibinfo {author} {\bibfnamefont {T.}~\bibnamefont
  {Bertalan}}, \bibinfo {author} {\bibfnamefont {F.}~\bibnamefont {Dietrich}},
  \bibinfo {author} {\bibfnamefont {I.}~\bibnamefont {Mezi{\'{c}}}}, \ and\
  \bibinfo {author} {\bibfnamefont {I.~G.}\ \bibnamefont {Kevrekidis}},\
  }\bibfield  {title} {\enquote {\bibinfo {title} {{On learning Hamiltonian
  systems from data}},}\ }\href {\doibase 10.1063/1.5128231} {\bibfield
  {journal} {\bibinfo  {journal} {Chaos: An Interdisciplinary Journal of
  Nonlinear Science}\ }\textbf {\bibinfo {volume} {29}},\ \bibinfo {pages}
  {121107} (\bibinfo {year} {2019})},\ \Eprint
  {http://arxiv.org/abs/1907.12715} {arXiv:1907.12715} \BibitemShut {NoStop}%
\bibitem [{\citenamefont {Greydanus}, \citenamefont {Dzamba},\ and\
  \citenamefont {Yosinski}(2019)}]{greydanus2019hamiltonian}%
  \BibitemOpen
  \bibfield  {author} {\bibinfo {author} {\bibfnamefont {S.}~\bibnamefont
  {Greydanus}}, \bibinfo {author} {\bibfnamefont {M.}~\bibnamefont {Dzamba}}, \
  and\ \bibinfo {author} {\bibfnamefont {J.}~\bibnamefont {Yosinski}},\
  }\href@noop {} {\enquote {\bibinfo {title} {Hamiltonian neural networks},}\ }
  (\bibinfo {year} {2019}),\ \Eprint {http://arxiv.org/abs/1906.01563}
  {arXiv:1906.01563 [cs.NE]} \BibitemShut {NoStop}%
\bibitem [{\citenamefont {Chen}\ \emph {et~al.}(2019)\citenamefont {Chen},
  \citenamefont {Zhang}, \citenamefont {Arjovsky},\ and\ \citenamefont
  {Bottou}}]{chen2019symplectic}%
  \BibitemOpen
  \bibfield  {author} {\bibinfo {author} {\bibfnamefont {Z.}~\bibnamefont
  {Chen}}, \bibinfo {author} {\bibfnamefont {J.}~\bibnamefont {Zhang}},
  \bibinfo {author} {\bibfnamefont {M.}~\bibnamefont {Arjovsky}}, \ and\
  \bibinfo {author} {\bibfnamefont {L.}~\bibnamefont {Bottou}},\ }\href@noop {}
  {\enquote {\bibinfo {title} {Symplectic recurrent neural networks},}\ }
  (\bibinfo {year} {2019}),\ \Eprint {http://arxiv.org/abs/1909.13334}
  {arXiv:1909.13334 [cs.LG]} \BibitemShut {NoStop}%
\bibitem [{\citenamefont {Burby}, \citenamefont {Tang},\ and\ \citenamefont
  {Maulik}(2020)}]{burby2020fast}%
  \BibitemOpen
  \bibfield  {author} {\bibinfo {author} {\bibfnamefont {J.~W.}\ \bibnamefont
  {Burby}}, \bibinfo {author} {\bibfnamefont {Q.}~\bibnamefont {Tang}}, \ and\
  \bibinfo {author} {\bibfnamefont {R.}~\bibnamefont {Maulik}},\ }\href@noop {}
  {\enquote {\bibinfo {title} {Fast neural poincaré maps for toroidal magnetic
  fields},}\ } (\bibinfo {year} {2020}),\ \Eprint
  {http://arxiv.org/abs/2007.04496} {arXiv:2007.04496 [physics.plasm-ph]}
  \BibitemShut {NoStop}%
\bibitem [{\citenamefont {Toth}\ \emph {et~al.}(2019)\citenamefont {Toth},
  \citenamefont {Rezende}, \citenamefont {Jaegle}, \citenamefont {Racanière},
  \citenamefont {Botev},\ and\ \citenamefont {Higgins}}]{toth2019hamiltonian}%
  \BibitemOpen
  \bibfield  {author} {\bibinfo {author} {\bibfnamefont {P.}~\bibnamefont
  {Toth}}, \bibinfo {author} {\bibfnamefont {D.~J.}\ \bibnamefont {Rezende}},
  \bibinfo {author} {\bibfnamefont {A.}~\bibnamefont {Jaegle}}, \bibinfo
  {author} {\bibfnamefont {S.}~\bibnamefont {Racanière}}, \bibinfo {author}
  {\bibfnamefont {A.}~\bibnamefont {Botev}}, \ and\ \bibinfo {author}
  {\bibfnamefont {I.}~\bibnamefont {Higgins}},\ }\href@noop {} {\enquote
  {\bibinfo {title} {Hamiltonian generative networks},}\ } (\bibinfo {year}
  {2019}),\ \Eprint {http://arxiv.org/abs/1909.13789} {arXiv:1909.13789
  [cs.LG]} \BibitemShut {NoStop}%
\bibitem [{\citenamefont {Rasmussen}\ and\ \citenamefont
  {Williams}(2005)}]{rasmussen2005}%
  \BibitemOpen
  \bibfield  {author} {\bibinfo {author} {\bibfnamefont {C.~E.}\ \bibnamefont
  {Rasmussen}}\ and\ \bibinfo {author} {\bibfnamefont {C.~K.~I.}\ \bibnamefont
  {Williams}},\ }\href@noop {} {\emph {\bibinfo {title} {Gaussian Processes for
  Machine Learning (Adaptive Computation and Machine Learning)}}}\ (\bibinfo
  {publisher} {The MIT Press},\ \bibinfo {year} {2005})\BibitemShut {NoStop}%
\bibitem [{\citenamefont {Solak}\ \emph {et~al.}(2003)\citenamefont {Solak},
  \citenamefont {Murray-smith}, \citenamefont {Leithead}, \citenamefont
  {Leith},\ and\ \citenamefont {Rasmussen}}]{solak2003}%
  \BibitemOpen
  \bibfield  {author} {\bibinfo {author} {\bibfnamefont {E.}~\bibnamefont
  {Solak}}, \bibinfo {author} {\bibfnamefont {R.}~\bibnamefont {Murray-smith}},
  \bibinfo {author} {\bibfnamefont {W.~E.}\ \bibnamefont {Leithead}}, \bibinfo
  {author} {\bibfnamefont {D.~J.}\ \bibnamefont {Leith}}, \ and\ \bibinfo
  {author} {\bibfnamefont {C.~E.}\ \bibnamefont {Rasmussen}},\ }\bibfield
  {title} {\enquote {\bibinfo {title} {Derivative observations in {Gaussian}
  process models of dynamic systems},}\ }in\ \href
  {http://papers.nips.cc/paper/2287-derivative-observations-in-gaussian-process-models-of-dynamic-systems.pdf}
  {\emph {\bibinfo {booktitle} {Advances in Neural Information Processing
  Systems 15}}},\ \bibinfo {editor} {edited by\ \bibinfo {editor}
  {\bibfnamefont {S.}~\bibnamefont {Becker}}, \bibinfo {editor} {\bibfnamefont
  {S.}~\bibnamefont {Thrun}}, \ and\ \bibinfo {editor} {\bibfnamefont
  {K.}~\bibnamefont {Obermayer}}}\ (\bibinfo  {publisher} {MIT Press},\
  \bibinfo {year} {2003})\ pp.\ \bibinfo {pages} {1057--1064}\BibitemShut
  {NoStop}%
\bibitem [{\citenamefont {Eriksson}\ \emph {et~al.}(2018)\citenamefont
  {Eriksson}, \citenamefont {Lee}, \citenamefont {Dong}, \citenamefont
  {Bindel},\ and\ \citenamefont {Wilson}}]{eriksson2018}%
  \BibitemOpen
  \bibfield  {author} {\bibinfo {author} {\bibfnamefont {D.}~\bibnamefont
  {Eriksson}}, \bibinfo {author} {\bibfnamefont {E.}~\bibnamefont {Lee}},
  \bibinfo {author} {\bibfnamefont {K.}~\bibnamefont {Dong}}, \bibinfo {author}
  {\bibfnamefont {D.}~\bibnamefont {Bindel}}, \ and\ \bibinfo {author}
  {\bibfnamefont {A.}~\bibnamefont {Wilson}},\ }\bibfield  {title} {\enquote
  {\bibinfo {title} {Scaling {Gaussian} process regression with derivatives},}\
  }\href@noop {} {\bibfield  {journal} {\bibinfo  {journal} {Advances in Neural
  Information Processing Systems}\ }\textbf {\bibinfo {volume}
  {2018-December}},\ \bibinfo {pages} {6867--6877} (\bibinfo {year} {2018})},\
  \bibinfo {note} {32nd Conference on Neural Information Processing Systems,
  NeurIPS 2018 ; Conference date: 02-12-2018 Through 08-12-2018}\BibitemShut
  {NoStop}%
\bibitem [{\citenamefont {O’Hagan}(1992)}]{ohagan1992}%
  \BibitemOpen
  \bibfield  {author} {\bibinfo {author} {\bibfnamefont {A.}~\bibnamefont
  {O’Hagan}},\ }\bibfield  {title} {\enquote {\bibinfo {title} {Some
  {Bayesian} numerical analysis},}\ }\href@noop {} {\bibfield  {journal}
  {\bibinfo  {journal} {Bayesian Statistics}\ }\textbf {\bibinfo {volume}
  {4}},\ \bibinfo {pages} {4--2} (\bibinfo {year} {1992})}\BibitemShut
  {NoStop}%
\bibitem [{\citenamefont {Álvarez}, \citenamefont {Rosasco},\ and\
  \citenamefont {Lawrence}(2012)}]{alvarez2012}%
  \BibitemOpen
  \bibfield  {author} {\bibinfo {author} {\bibfnamefont {M.~A.}\ \bibnamefont
  {Álvarez}}, \bibinfo {author} {\bibfnamefont {L.}~\bibnamefont {Rosasco}}, \
  and\ \bibinfo {author} {\bibfnamefont {N.~D.}\ \bibnamefont {Lawrence}},\
  }\bibfield  {title} {\enquote {\bibinfo {title} {Kernels for vector-valued
  functions: A review},}\ }\href {\doibase 10.1561/2200000036} {\bibfield
  {journal} {\bibinfo  {journal} {Foundations and Trends in Machine Learning}\
  }\textbf {\bibinfo {volume} {4}},\ \bibinfo {pages} {195--266} (\bibinfo
  {year} {2012})}\BibitemShut {NoStop}%
\bibitem [{\citenamefont {Seber}\ and\ \citenamefont
  {Lee}(2012)}]{seber2012linear}%
  \BibitemOpen
  \bibfield  {author} {\bibinfo {author} {\bibfnamefont {G.}~\bibnamefont
  {Seber}}\ and\ \bibinfo {author} {\bibfnamefont {A.}~\bibnamefont {Lee}},\
  }\href {https://books.google.de/books?id=X2Y6OkXl8ysC} {\emph {\bibinfo
  {title} {Linear Regression Analysis}}},\ Wiley Series in Probability and
  Statistics\ (\bibinfo  {publisher} {Wiley},\ \bibinfo {year}
  {2012})\BibitemShut {NoStop}%
\bibitem [{\citenamefont {Fasshauer}(1997)}]{fasshauer1996}%
  \BibitemOpen
  \bibfield  {author} {\bibinfo {author} {\bibfnamefont {G.~E.}\ \bibnamefont
  {Fasshauer}},\ }\bibfield  {title} {\enquote {\bibinfo {title} {Solving
  partial differential equations by collocation with radial basis functions},}\
  }in\ \href@noop {} {\emph {\bibinfo {booktitle} {In: Surface Fitting and
  Multiresolution Methods A. Le M'ehaut'e, C. Rabut and L.L. Schumaker (eds.),
  Vanderbilt}}}\ (\bibinfo  {publisher} {University Press},\ \bibinfo {year}
  {1997})\ pp.\ \bibinfo {pages} {131--138}\BibitemShut {NoStop}%
\bibitem [{\citenamefont {Kansa}(1990)}]{kansa1990}%
  \BibitemOpen
  \bibfield  {author} {\bibinfo {author} {\bibfnamefont {E.}~\bibnamefont
  {Kansa}},\ }\bibfield  {title} {\enquote {\bibinfo {title} {Multiquadrics—a
  scattered data approximation scheme with applications to computational
  fluid-dynamics—ii solutions to parabolic, hyperbolic and elliptic partial
  differential equations},}\ }\href {\doibase
  https://doi.org/10.1016/0898-1221(90)90271-K} {\bibfield  {journal} {\bibinfo
   {journal} {Computers \& Mathematics with Applications}\ }\textbf {\bibinfo
  {volume} {19}},\ \bibinfo {pages} {147 -- 161} (\bibinfo {year}
  {1990})}\BibitemShut {NoStop}%
\bibitem [{\citenamefont {Raissi}, \citenamefont {Perdikaris},\ and\
  \citenamefont {Karniadakis}(2017)}]{raissi2017}%
  \BibitemOpen
  \bibfield  {author} {\bibinfo {author} {\bibfnamefont {M.}~\bibnamefont
  {Raissi}}, \bibinfo {author} {\bibfnamefont {P.}~\bibnamefont {Perdikaris}},
  \ and\ \bibinfo {author} {\bibfnamefont {G.~E.}\ \bibnamefont
  {Karniadakis}},\ }\bibfield  {title} {\enquote {\bibinfo {title} {Inferring
  solutions of differential equations using noisy multi-fidelity data},}\
  }\href {\doibase https://doi.org/10.1016/j.jcp.2017.01.060} {\bibfield
  {journal} {\bibinfo  {journal} {Journal of Computational Physics}\ }\textbf
  {\bibinfo {volume} {335}},\ \bibinfo {pages} {736 -- 746} (\bibinfo {year}
  {2017})}\BibitemShut {NoStop}%
\bibitem [{\citenamefont {Albert}\ and\ \citenamefont
  {Rath}(2020)}]{albert2020}%
  \BibitemOpen
  \bibfield  {author} {\bibinfo {author} {\bibfnamefont {C.~G.}\ \bibnamefont
  {Albert}}\ and\ \bibinfo {author} {\bibfnamefont {K.}~\bibnamefont {Rath}},\
  }\bibfield  {title} {\enquote {\bibinfo {title} {Gaussian process regression
  for data fulfilling linear differential equations with localized sources.}}\
  }\href {https://doi.org/10.3390/e22020152} {\bibfield  {journal} {\bibinfo
  {journal} {Entropy}\ }\textbf {\bibinfo {volume} {22}},\ \bibinfo {pages}
  {152} (\bibinfo {year} {2020})}\BibitemShut {NoStop}%
\bibitem [{\citenamefont {Zotos}(2017)}]{Zotos}%
  \BibitemOpen
  \bibfield  {author} {\bibinfo {author} {\bibfnamefont {E.~E.}\ \bibnamefont
  {Zotos}},\ }\bibfield  {title} {\enquote {\bibinfo {title} {{An overview of
  the escape dynamics in the H{\'{e}}non-Heiles Hamiltonian system}},}\ }\href
  {\doibase 10.1007/s11012-017-0647-8} {\bibfield  {journal} {\bibinfo
  {journal} {Meccanica}\ }\textbf {\bibinfo {volume} {52}} (\bibinfo {year}
  {2017}),\ 10.1007/s11012-017-0647-8}\BibitemShut {NoStop}%
\bibitem [{\citenamefont {Berndt}, \citenamefont {Klucznik},\ and\
  \citenamefont {Society}(2001)}]{berndt2001}%
  \BibitemOpen
  \bibfield  {author} {\bibinfo {author} {\bibfnamefont {R.}~\bibnamefont
  {Berndt}}, \bibinfo {author} {\bibfnamefont {M.}~\bibnamefont {Klucznik}}, \
  and\ \bibinfo {author} {\bibfnamefont {A.~M.}\ \bibnamefont {Society}},\
  }\href {https://books.google.de/books?id=5HbJJNYXoN0C} {\emph {\bibinfo
  {title} {An Introduction to Symplectic Geometry}}},\ Contemporary
  Mathematics\ (\bibinfo  {publisher} {American Mathematical Society},\
  \bibinfo {year} {2001})\BibitemShut {NoStop}%
\bibitem [{\citenamefont {José}\ and\ \citenamefont
  {Saletan}(1998)}]{saletan1998}%
  \BibitemOpen
  \bibfield  {author} {\bibinfo {author} {\bibfnamefont {J.~V.}\ \bibnamefont
  {José}}\ and\ \bibinfo {author} {\bibfnamefont {E.~J.}\ \bibnamefont
  {Saletan}},\ }\href {\doibase 10.1017/CBO9780511803772} {\emph {\bibinfo
  {title} {Classical Dynamics: A Contemporary Approach}}}\ (\bibinfo
  {publisher} {Cambridge University Press},\ \bibinfo {year}
  {1998})\BibitemShut {NoStop}%
\end{thebibliography}%

\end{document}